\documentclass[10pt]{article}
\usepackage{amssymb}
\usepackage{amsmath}
\usepackage{wasysym}
\usepackage{graphicx}
\begin{document}

\title{Lighthill equation for quantum liquids}

\author{C. Dedes\thanks{{c\_dedes@yahoo.com}}\\
Bradford College  \\
Great Horton Road, Bradford \\
West Yorkshire, BD7 1AY \\
United Kingdom \\}

 \maketitle

\begin{abstract}
A quantum version of the Lighthill equation that originated in the field of theoretical aeroacoustics is derived for the probability density of a superfluid starting from the time-dependent Gross-Pitaevskii equation. It involves a second-order time derivative and should be supplemented by two-time boundary conditions. Its physical implications are discussed in relation to the quantum equilibrium hypothesis and the general applicability of Born's rule.

\end{abstract}
The hydrodynamic formulation of quantum theory has a long history, starting with Madelung \cite{Madelung} almost immediately after the discovery of quantum mechanics and coming to maturity with the work of Takabayasi and others \cite{Takabayasi,Harvey, Wilhelm, Wong, Birula,Holland, Wyatt, Bush}. According to this picture there is a close analogy between motion of material and probability fluids which is made apparent by two continuity equations of the Navier-Stokes type for the probability density and the momentum of this probability field. This indicates that quantum phenomena may be modeled in a hydrodynamic framework. It should be remembered on the other hand that the material fluids exist in the three dimensional space but a many particle wavefunction in an abstract configuration space. Furthermore, the stress tensor in the case of a probability fluid has distinctive quantum features. This alternative formulation of quantum mechanics, which shares many similarities with the pilot wave theory \cite{Wyatt}, did not prove especially popular among physicists. Yet the hydrodynamical picture is particular successful when it comes to the description of ultra cold atoms and weak interacting quantum gases \cite{Primer, Spiegel} even though the physical content is rather different in that occasion, since the macroscopic single-particle complex wavefunction that describes a condensate is not a quantum state and the equation of motion is non-linear. The time-dependent Gross-Pitaevskii equation that governs the time evolution of a quantum superfluid is written as

\begin{equation}
i\hbar\frac{\partial \Psi}{\partial t} = 
 \left(-\frac{\hbar ^{2}}{2m}\nabla ^{2}+V(\mathbf{r},t)+g|\Psi|^{2}\right)\Psi , 
\end{equation}

\noindent
where g is a coupling constant, $V(\mathbf{r},t)$ the external potential and $\Psi$ a single-particle wavefunction.

A brief digression is befitting here in order to provide some motivation for what will follow. An initial observation is that the non-linear wave equation (1) contains a first order derivative in time so it resembles a diffusion equation in that respect which means that $\Psi(t)\neq \Psi(-t)$. According to the Wigner prescription $\Psi(t)\rightarrow \Psi ^{*}(-t)$ it follows that the particle density is time symmetric $n(t)=n(-t)$. It seems reasonable then to investigate if we could derive a manifestly time-invariant evolution equation for $n(t)$, possibly second order in time. In other words we exploit the Gross-Pitaevskii equation for the macroscopic wavefunction to deduce an equation of motion for the probability density. To achieve this we proceed by introducing a Madelung polar transformation

\begin{equation}
\Psi=\sqrt{n}e^{iS/\hbar},
\end{equation}

\noindent
where $S$ is the multivalued phase of the wavefunction and the velocity field coupled to a vector potential is expressed as

\begin{equation}
\mathbf {v}=\frac{\nabla S-q\mathbf{A}}{m}.
\end{equation}

\noindent
After separating real and imaginary parts two formulas are obtained. By equating the imaginary parts we arrive at a probability density continuity equation which is written in index notation, where summation over repeated indices is implied, as

\begin{equation}
 \frac{\partial n}{\partial t}+\frac{\partial (nv_{j})}{\partial x_{j}}=0. 
\end{equation}

\noindent
The Hamilton-Jacobi equation is obtained from (1) through (2) by equating its real parts. 
 
 \begin{equation}
     \frac{\partial S}{\partial t}=-\frac{1}{2m}\left({\nabla S}\right)^{2}-V(\mathbf{r},t)-Q-gn.
 \end{equation}
 
 \noindent
  In the above expression $Q=-\frac{\hbar ^{2}}{2m}\frac{\nabla ^{2}\sqrt{n}}{\sqrt{n}}$ is the Bohm quantum potential.Taking the divergence gives the second hydrodynamic equation 
  
 \begin{equation}
    \frac{\partial (n \mathbf{v})}{\partial t}+\nabla \cdot ( n \mathbf{v}\mathbf{v})=-\nabla (V+Q+gn),
\end{equation}

\noindent
where the quantum force is the divergence of the sum of both classical and quantum potentials. It is possible to write the above in a tensor form as in \cite{Primer}

\begin{equation}
  \frac{\partial (n v_{i})}{\partial t}+\frac{\partial (n u_{i}v_{j})}{\partial x_{j}}=-\frac{n}{m} \frac{\partial V}{\partial x_{i}}-\frac{\partial\Pi_{ij}}{\partial x_{j}}-\frac{g}{2m}\frac{\partial (n^{2})}{\partial x_{i}}.
\end{equation}

\noindent
The probability quantum stress tensor is written as

\begin{equation}
  \Pi _{ij}=\frac{\hbar ^{2}}{4m^{2}}\left(\nabla_{i}\nabla _{j}n-\nabla_{i}n\nabla _{j}lnn \right).
  \end{equation}

 \noindent
Notice that the stress tensor does not depend on the velocity field but only on the density of the probability fluid. It is a second rank tensor with nine-components for a single particle in the three dimensional space, whereas for $N$ particles the number of terms is $N^{2}$. When $g=0$ we recover the familiar hydrodynamic equations that follow from the single particle Schr\"{o}dinger equation \cite{Holland, Harvey, Wilhelm}. Since we have performed differentiation the hydrodynamical equations we obtained are not equivalent to the Gross-Pitaevskii unless we impose certain constraints on the line integral of the velocity field (see the relevant discussion in \cite{Birula,Wallstrom} for the Schr\"{o}dinger equation). As we do not aim to discard the wave equation or to recover the wavefunction equation from the hydrodynamic ones, this does not affect our scope.

It is possible to rearrange and merge (4) and (7) into a single equation in a fashion similar to Lighthill's who long ago derived an evolution equation for the generation of aerodynamic sound \cite{Lighthill,Howe,Mattei}. Differentiating  over time (4) and taking the divergence of (7) and then subtracting the two resulting expressions we manage to eliminate the terms associated with the momentum of the probability fluid. To make more explicit the connection to the Lighthill equation we further subtract from both sides a $c_{0} ^{2} \nabla^{2}n$, where $c_{0}$ an arbitrary reference speed, and arrive at the following time-invariant formula

  \begin{equation}
   \frac{\partial ^{2}n}{\partial t ^{2}}-c_{0} ^{2} \nabla^{2}n =\frac{\partial ^{2}}{\partial x_{i}\partial x_{j}}\left[n v_{i}v_{j}-n\left( \frac{V}{m}+c_{0}^{2}\right)\delta _{ij}+\Pi_{ij}-\frac{gn^{2}}{2m}\right],
   \end{equation}

 \noindent
 where the expression inside brackets in the right hand side is a Lighthill source tensor which includes a quantum component with highly non-classical properties. Hence we have obtained the equation we sought for $n(t)$. This is our central result, an exact, non-linear tensor equation for the probability density that is driven by the velocity field governed by (1). We notice that the particle density appears also in the right hand side in a non-linear manner through the quantum stress tensor and other terms. Formally the solution can be written in integral form as

  \begin{equation}
       n(\mathbf{x},t) = \frac{1}{4\pi c_{0} ^{2}}\frac{\partial ^{2}}{\partial x_{i}\partial x_{j}}\int _{-\infty}^{+\infty} \frac{T_{ij}\left(\mathbf{y},t-\frac{|\mathbf{x}-\mathbf{y}|}{c_{0}}\right)}{|\mathbf{x}-\mathbf{y}|}d^{3}\mathbf{y},
   \end{equation}

 \noindent
 where $r=|\mathbf{x}-\mathbf{y}|$ the distance between reception and source and $T_{ij}$ the Lighthill tensor mentioned earlier. Notice, that we have included only the retarded solution but advanced ones may be included too when dealing with appropriately pre- and post-selected quantum ensembles (in that case we require two-time boundary conditions \cite{Schulman} for the particle density and its first time derivative). Due to the non-linear character of the source tensor in (9), even when $g=0$ which means that the driving equation (1) is linear, makes this equation difficult to solve for practical purposes yet it is proposed here for the conceptual insight it may offer to the study of foundational quantum mechanical theory. Some remarks should be included for the case with $g=0$. Firstly, a solution $\Psi$ multiplied by a complex number $w$ yields a new solution $w\Psi$. This illustrates the homogeneity property which is reflected into (9) when we multiply a solution $n$ by a factor $|w|^{2}$. Secondly, since the evolution equation for $n$ is non-linear, a superposition of number densities does not constitute a new solution. Yet, a superposition of wavefunction solutions satisfies both the linear Schr\"{o}dinger equation and (9) provided that $g=0$.

 An equation second order in time derived from linearized hydrodynamic equations is given in \cite{Nore, Salazar}. We instead linearize (9) by letting $n\rightarrow n_{0}+\delta n$ and let $g=0$ ($n_{0}=|\Psi|^{2}$ is an equilibrium solution as will be explained in the final paragraph and $\delta n$ a small perturbation). Keeping first order perturbation terms yields

    \begin{equation}
   \frac{\partial ^{2}\delta n}{\partial t ^{2}}-c_{0} ^{2} \nabla^{2}\delta n =\frac{\partial ^{2} }{\partial x_{i}\partial x_{j}}\left\{  \left[v_{i}v_{j}-\left( \frac{V}{m}+c_{0}^{2}\right)\delta _{ij}\right]\delta n+\delta \Pi_{ij}\right\},
   \end{equation}
   
   \noindent
  with a stress tensor perturbation

    \begin{equation}
    \delta \Pi_{ij}=\frac{\hbar ^{2}}{4m^{2}}\left[\nabla _{i}\nabla _{j}-\nabla_{i}lnn_{0} \nabla_{j}-\nabla _{j}lnn_{0} \nabla_{i} +\nabla _{j}lnn_{0}\nabla _{i}lnn_{0}\right]\delta n. 
   \end{equation}

A word of caution is appropriate at this point. The Gross-Pitaevskii equation (or its linear counterpart when $g=0$) is not to be replaced by (9). Neither do we seek to derive the Schr\"{o}dinger equation (linear or non-linear)  from it. This kind of approach is misguided as both (1) and (9) should be used in conjunction. From a more philosophical perspective we might say that the wavefunction $\Psi$ belongs to the realm of quantum potentialities and is dictated by (1). On the other hand its absolute square which is directly measurable and seems to lay in a higher plane of ontological density is governed by (9). Yet, it is the wavefunction evolution equation that guides the probability density through the velocity field. In the de Broglie-Bohm picture the Schr\"{o}dinger equation guides the unobservable quantum trajectories through the same velocity field and here we have moved one step further. Naturally, a question arises about the solutions that (9) admits. It is evident that $n=|\Psi|^{2}$ satisfies the quantum Lighthill equation but this does not necessarily exhaust every conceivable solution, it is possible that there are distinct solutions of the latter such that $n \neq |\Psi|^{2}$. In that case we would have deviations from Born's rule and the quantum equilibrium hypothesis would not hold. This possibility has been argued by various authors \cite{Cushing} using different kinds of argumentation from the one presented here. It should be remembered nevertheless that not every solution of an equation of motion, even an exact one, must inevitably be instantiated in nature since it could be unstable \cite{Landau} and sensitive even to slight disturbances to remain observable.

\end{document}